\RequirePackage{fix-cm}
\documentclass[smallextended]{svjour3}       
\smartqed  
\usepackage{graphicx}

\usepackage{amsmath}
\usepackage{amssymb}
\usepackage{hyperref}
\usepackage[caption=false]{subfig}

\usepackage[final]{changes}

\usepackage{lipsum}
\definechangesauthor[color=blue]{.}
\colorlet{Changes@Color}{red}
\setremarkmarkup{(#2)}

\begin{document}

\title{Landau quasiparticles in weak power-law liquids
}


\author{Joshuah T. Heath
}


\institute{J.T. Heath \at 
	 Physics Department \\ Boston  College \\ 140 Commonwealth Avenue \\ Chestnut Hill, MA 02467 \\
	 Tel.:+802-922-1291 \\
              \email{heathjo@bc.edu}           
           \and
}

\date{Received: date / Accepted: date}

\maketitle

\begin{abstract}
The failure of Landau-Fermi liquid theory is often considered a telltale sign of universal, scale-invariant behavior in the emergent field theory of interacting fermions. Nevertheless, there exist borderline cases where weak scale invariance coupled with particle-hole asymmetry can coexist with the Landau quasiparticle paradigm. In this letter, I show explicitly that a Landau-Fermi liquid can exist for weak power-law scaling of the retarded Green's function. Such an exotic variant of the traditional Fermi liquid is shown to always be incompatible with Luttinger's theorem for any non-trivial scaling. This result yields evidence for a Fermi liquid-like ground state in the high-field, underdoped pseudogap phase of the high-temperature cuprate superconductors.
\end{abstract}


\newpage

\noindent {\it I. {Introduction.}}--In recent years, the study of non-Fermi liquids has been dominated by quantum critical phenomena \cite{Hertz,Millis,Schofield,Sachdev_book,Su}. In particular, the transport quantities of the optimally-doped cuprate superconductors are now known to exhibit quantum critical scaling, with the the electrical resistivity \cite{Ando,Hussey,Cooper}, the Hall angle \cite{Chien,Tyler}, and the Lorentz ratio \cite{Zhang} all exhibiting temperature dependencies inconsistent with super-Planckian lifetimes of the charge carriers \cite{Hartnoll,Karch}. Nevertheless, the underlying microscopic origin of such behavior is still a matter of debate, with proposals ranging from coupled SYK islands \cite{Patel1,Patel2,Patel3} to an exotic phase of matter which lacks charge quantization \cite{Phillips1,Phillips2}. 

Mirroring the microscopic underpinnings of conventional Landau-Fermi liquids \cite{Landau1,Migdal,Migdal2,Gorkov_book}, Anderson first proposed that the unusual normal phase of the optimally-doped high-$T_c$ cuprates can be explained via a "hidden Fermi liquid theory" \cite{Anderson1,Anderson2}.
The Fourier-transformed Green's function of such a system can be written as a Landau-like propagator raised to a power proportional to the sum of phase shifts over all scattering channels, reducing the quasiparticle pole to a branch cut. 
Such "power-law liquids"\footnote{Although power-law liquids were originally defined as a system characterized by a power-law scaling of the self energy \cite{Reber}, I take the terminology to mean some general scaling of the Green's function itself \cite{Phillips4} throughout this paper.} have recently been shown to violate the Luttinger sum rule \cite{Phillips4}, and interact with conventional electrons in such a way that may lead to a power-law scaling of the imaginary self energy similar to what is seen in ARPES experiments on Bi$_2$Sr$_{2}$CaCu$_2$O$_{8+\delta}$ in the underdoped pseudogap phase \cite{Reber,Phillips3,Phillips5}. 
Indeed, such power-law behavior is often taken as the hallmark of an "unparticle"-like phase, where scale invariance naturally leads to a power-law Green's function and the Standard Model notion of an independent particle breaks down \cite{Georgi,Georgi2}. Nevertheless, there have been several cases in the past few years that have shown that Landau-Fermi liquid theory persists in close proximity to a scale-invariant quantum critical point \cite{Gochan,Wolfle1,Wolfle2,Wolfle3}, contradicting the underlying predictions of an unparticle-like condensate.

In this letter, I will show that power-law scaling of the electronic Green's function does not necessarily imply a non-trivial IR fixed point in the many-body field theory. 
Specifically, in dimensions $d\ge 2$, the spectral density exhibits a clearly defined peak and weight for Green's functions raised to reasonably small positive powers. The existence of a finite quasiparticle weight is shown to always be incompatible with Luttinger's theorem for all non-zero powers, supporting the claim that the underdoped pseudogap phase of the cuprates at high magnetic fields \cite{Doiron} may be a Luttinger's theorem-violating Landau-Fermi liquid \cite{Wen2} or "fractionalized" Fermi liquid \cite{Sachdev1,Punk,Sachdev2,Sachdev3}.
\\\\
{\it II. Spectral density of power-law Green's functions.}-- I consider Green's functions for interacting fermions raised to some power $\xi \in \mathbb{R}$:
%
\begin{align}
G_{\xi}({\bf k},\,\omega)=\frac{1}{(\omega-\epsilon_{\bf k}-\Sigma({\bf k},\omega)+i\delta)^{1+\xi}}\label{eq1}
\end{align}
where the self energy $\Sigma({\bf k},\,\omega)$ is given by
$\Sigma({\bf k},\,\omega)=(G^{(0)}_0)^{-1}-G_0^{-1}$. If $\xi=0$, we obtain the well-known propagator of the dressed Landau quasiparticle, with the presence of a finite pole dictated by the frequency-dependence of the real part of the self energy \cite{Migdal,Bedell1,Heath1}.

 If $\xi\not=0$, it is often claimed that the resultant branch cut kills off any well-defined quasiparticle, but this claim is misguided. The oft-quoted example of such a power-law non-Fermi liquid is the 1D Tomonaga-Luttinger model \cite{Haldane1,Wen_Luttinger,Voit,Sierra_book}, which is described by a propagator of the form

\begin{align}
G_{\pm}({\bf k},\,\omega)=\frac{(\widetilde{v}_F^2 k^2-\omega^2)^{s^2/2-1}}{|(\pm\widetilde{v}_F k-\omega)(\pm v_F k-\omega)|^{1/2}}
\end{align}
where $\widetilde{v}_F$ and $s$ are dependent on the interaction and $\pm$ denote right- and left-handed fermions, respectively. 
However, it is crucial to note that branch cut singularities for the Tomonaga-Luttinger system exist on either side of the Fermi point by virtue of the $g$-ology construction of the 1D Hamiltonian \cite{Bedell1}.
For higher-dimensional systems characterized by Green's functions such as Eqn. \eqref{eq1}, 
we can simply close the contour on the opposite side of the real-frequency axis to preserve the pole structure \cite{Bedell1}, {assuming that the branch cut does not include the singularity of the Green's function (i.e., the branch cut does not include the Fermi point/surface). Physically, when the branch cut replaces the pole, the bare electron distribution function is characterized by an infinite slope as opposed to a finite discontinuity, resulting in either Luttinger liquid or "marginal Fermi liquid" behavior \cite{Solyom,Ruckenstein}. In the case where branch cuts coexist with poles, however, the Fermi liquid picture has been shown to remain, albeit with non-analyticities in the thermodynamic potential \cite{Finkle}. In a similar fashion,} a higher-dimensional system composed of Tomonaga-Luttinger "threads" coupled via a Coulomb interaction display power-law behavior in the electron propagator, yet exhibits a finite discontinuity in the momentum dispersion unseen in the "true" 1D model \cite{Larkin}. It should therefore be apparent that the breakdown of Landau-Fermi liquid theory in one dimension (and fractional dimensions $1<d<2$ \cite{Wen1}) is an inherent consequence of dimensional reduction, and not from {the appearance of a branch cut from} the power-law nature of the propagator. 

To investigate the possibility of Landau-like excitations in the power-law liquid, I solve for the spectral density $A({\bf k},\,\omega)\sim \Im G_\xi ({\bf k},\,\omega)$ resulting from Eqn. \eqref{eq1}. {To do this, I simplify the power-law Green's function via the following:}

\begin{align}
{G(k,\,\omega)}&{=\frac{1}{(\omega -\epsilon_k-\Sigma ({\bf k},\,\omega)+i\delta)^{1+\xi}}}\notag\\
&{=\sum_{n=0}^\infty \frac{(-\xi)^n}{n!}\frac{(\log(\omega-\epsilon_k-\Sigma({\bf k},\,\omega)+i\delta))^n}{\omega-\epsilon_k-\Sigma ({\bf k},\,\omega)+i\delta}}\label{Eqn3}
\end{align}
{This allows us to write the spectral function in the form}

\begin{align}
{A\sim}&{ -\frac{1}{\pi}\frac{\Im \Sigma({\bf k},\,\omega)}{(x-\Re \Sigma ({\bf k},\,\omega))^2+\Im \Sigma({\bf k},\,\omega)^2}}\notag\\
&{\times\sum_{n=0}^\infty \frac{(-\xi)^n}{n!}\left\{
r^n(\cos(n\theta)+\frac{x-\Re\Sigma({\bf k},\,\omega)}{\Im \Sigma({\bf k},\,\omega)}\sin(n\theta))
\right\}}\label{A}
\end{align}
{where $x\equiv \omega-\epsilon_{\bf k}$ and $r$ and $\theta$ are the modulus and phase of the complex logarithm in Eqn. \eqref{Eqn3}, respectively. We can simplify further by performing each sum separately and recalling the phase $\phi_0(\omega)$ of the retarded Green's function for $\xi=0$:}

\begin{align}
{\tan \phi_0(\omega)=\frac{\Im(G_0(k,\,\omega))}{\Re(G_0(k,\,\omega))}=\frac{\Im\Sigma ({\bf k},\,\omega)}{x-\Re\Sigma ({\bf k},\,\omega)}}
\end{align}
{The behavior of such a phase is central to the discussion of Luttinger's theorem in these power-law liquids, which is the topic of the next section.}

{The above allows us to write the spectral function in the form}

\begin{align}
{
\frac{A_\xi(k,\,\omega)}{A_0(k,\,\omega)}=\frac{
\cos [\xi \phi_0(\omega)]-\cot[\phi_0(\omega)]\sin[\xi \phi_0(\omega)]
}{((x-\Re\Sigma({\bf k},\,\omega))^2+\Im\Sigma({\bf k},\,\omega)^2)^{\xi/2}
}}\label{ratio}
\end{align}
{where I take the notation $A_\xi({\bf k},\,\omega)$ to mean the spectral function at some non-zero power $\xi$ and $A_0({\bf k},\,\omega)$ to be the spectral function for trivial power $\xi=0$. Note that, as $\xi\rightarrow 0$, the right-hand side approaches unity, as expected.}

{With the spectral function written as Eqn. \eqref{ratio}, we can now write down an approximate form near the Fermi surface:}

\begin{align}
{A_{\xi}(k,\,\omega)}&{=A_0(k,\,\omega)\left\{
\frac{
\cos [\xi \phi_0(\omega)]-\cot[\phi_0(\omega)]\sin[\xi \phi_0(\omega)]
}{((x-\Sigma')^2+\Sigma''^2)^{\xi/2}
}
\right\}}\notag\\
&{\sim-\frac{1}{\pi} \frac{\Sigma'' Z_k^2}{\widetilde{x}^2+\Sigma''^2 Z_k^2}\left\{
Z_k^\xi
\frac{
\cos [\xi \phi_0(\omega)]-\cot[\phi_0(\omega)]\sin[\xi \phi_0(\omega)]
}{(\widetilde{x}^2+\Sigma''^2Z_k^2)^{\xi/2}
}
\right\}+A_{inc}}\notag\\
&{\sim\frac{\widetilde{Z}_k}{\pi} \frac{1/\tau}{(\widetilde{x}^2+(1/\tau)^2)^{1+\xi/2}} +A_{inc}\label{spectral}}
\end{align}
%
%
%
%
{where $A_{inc}$ is the incoherent part of the spectral weight, $\tau$ is the width of the spectral density for $\xi=0$, and} the "effective" quasiparticle weight is given by
\begin{align}
\widetilde{Z}_k=Z_k^{1+\xi} \bigg\{\cos[
\xi \phi_0(0)
]-\cot[\phi_0(0)]\sin[\xi\phi_0(0)]\bigg\}
\end{align}
{where $Z_k$ is the conventional quasiparticle weight for a Fermi liquid at $\xi=0$, given by} 
\begin{align}
{Z_k=\left(1-\frac{\partial \Re \Sigma({\bf k},\,\omega)}{\partial \omega}\bigg|_{\omega=0}\right)^{-1}}
\end{align}
{The above formulation of the power-law Green's function therefore allows us to recast this system in terms of a conventional, Fermi liquid-like Green's function with the power-law dependence buried in a re-scaled quasiparticle residue:}
\begin{align}
{G^R}&{=\int_{-\infty}^\infty d\omega' \frac{A_\xi({\bf k},\,\omega)}{\omega-\omega'+i\delta}}\notag\\
&{=\int_{-\infty}^\infty d\omega' \frac{1}{\omega-\omega'+i\delta}\left\{
 \frac{\widetilde{Z}_k}{\pi} \frac{1/\tau}{(\widetilde{x}^2+(1/\tau)^2)^{1+\xi/2}}+A_{inc}
\right\}}\notag\\
&{=\frac{\widetilde{Z}_k}{
\widetilde{x}-\frac{i}{\tau_k}
}+G^R_{inc}}
\end{align}
{The dependence on $\xi$ in the "effective" quasiparticle weight tells us that non-trivial power-law scaling of the Green's function scales the discontinuity of the electronic distribution function in a non-trivial way. Later on, we will see that a finite subset of powers $\xi$ will result in a stable Fermi liquid; i.e., a value $0<\widetilde{Z}_k\le 1$.}

{The final form of the spectral weight can be found by imposing the sum rule $\int_{-\infty}^\infty A({\bf k},\,\omega)=1$ on Eqn. \eqref{spectral}. This follows from the fact that $A({\bf k},\,\omega)$ can still be interpreted as probability density for non-trivial powers $\xi>-1$ (as previously shown in \cite{Phillips4}), and yields the appropriate normalization constant: }
%
%
\begin{align}
&A_\xi({\bf k},\,\omega)
=\widetilde{Z}_k\left\{
\frac{
\Gamma\left(1+\frac{\xi}{2}\right)
}{
\Gamma\left(
\frac{1+\xi}{2}
\right)\sqrt{\pi}
} \bigg[\frac{
1/\tau^{1+\xi}
}{
(\widetilde{x}^2+(1/\tau)^2)^{1+\xi/2}
}\bigg]
\right\}\label{Eqn6}
\end{align}
%

\noindent The normalization constant in Eqn. \eqref{Eqn6} was derived assuming that $\xi>-1$; otherwise, the integral fails to converge. This makes sense from a physical point of view by looking at the branch cut structure of Eqn. \eqref{eq1}: when $\xi \in \mathbb{R}/\mathbb{Z}$, the branch cut consists of a line of complex numbers $z\in (-\infty,\,0)$ for $\xi>-1$ and $z\in (-\infty,\,0]$ for $\xi<-1$. In the case of the latter, the branch cut includes the point of the Green's function singularity, resulting in a finite quasiparticle lifetime at the Fermi surface and a failure of the quasiparticle paradigm. More drastically, the regime of $\xi<-1$ results in an unphysical (i.e., negative) spectral weight, telling us that the proposal Eqn. \eqref{Eqn6} is incorrect and no Landau-like quasiparticle can survive for power-law scaling $\xi<-1$. {This should be apparent from the form of \eqref{Eqn3}, where the regime of $\xi<-1$ transforms the Fermi surface into a Luttinger surface (i.e., a surface of zeroes of the Green's function as opposed to the Green's function's inverse \cite{Dzyal}).}

\begin{figure*}
 \begin{center}{\hspace{0mm} \includegraphics[width=.9\columnwidth]{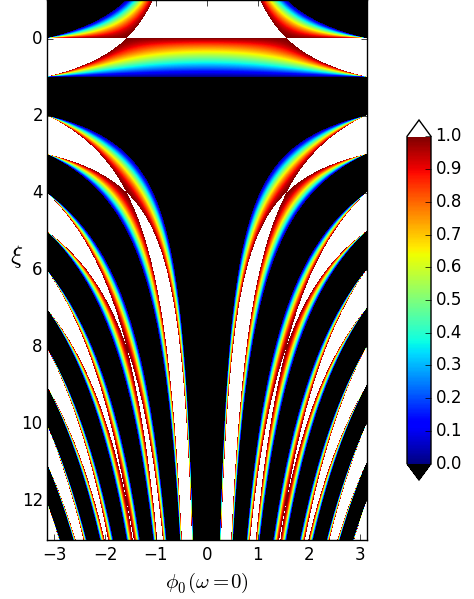} }\end{center}
\caption{A density profile of $\widetilde{Z}_k/Z_k^{\xi+1}$ plotted on a graph of the phase of the $\xi=0$ retarded Green's function $\phi_0(\omega=0)$ vs. the power $\xi$ of the power-law Green's function. The colour spectrum represents the regime where $0<\widetilde{Z}_k/Z_k^{\xi+1}\le 1$, illustrating where Landau-Fermi liquid theory will unanimously survive in the presence of non-trivial powers $\xi$. The white regions denote where $\widetilde{Z}_k/Z_k^{\xi+1}>1$, leading us to conclude that only weak Landau-Fermi liquids (i.e., $Z_k<<1$) can survive in this region. The black regions correspond to $\widetilde{Z}_k/Z_k^{\xi+1}<0$, and hence the complete breakdown of Landau-Fermi liquid theory. Note that a Landau-Fermi liquid-like quasiparticle weight is most stable for weak coupling $-1<\xi<1$, while a stable Fermi liquid for $\xi>1$ is only possible for certain non-trivial values of the phase $\phi_0(\omega=0)$.}
\label{fig1}
\end{figure*}%

For the case of $\xi>-1$, the spectral density Eqn. \eqref{Eqn6} displays a sharp peak not unlike what is seen in a traditional Landau-Fermi liquid. The corresponding effective lifetime $\widetilde{\tau}$ can be read off as the width of the Lorentzian:
\begin{align}
\widetilde{\tau}=\tau \sqrt{\pi}\frac{
\Gamma\left(
1+\frac{\xi}{2}
\right)
}{
\Gamma\left(
\frac{1+\xi}{2}
\right)
}
\end{align}
The above illustrates that the quasiparticle lifetime will remain finite as long as the power-law spectral function is of the form Eqn. \eqref{Eqn6}. Note that $\xi>0$ will result in an apparent increase in the quasiparticle lifetime, while $-1<\xi<0$ results in a reduction.

To gain more insight into the stability or instability of Landau-like excitations in power-law Green's functions, the ratio of the effective weights $\widetilde{Z}_k/Z_k^{\xi+1}$ is plotted in Fig. \ref{fig1} as a density plot of $\phi_0(\omega=0)$ vs. $\xi$. 
{The conventional result $\phi_0(\omega=0)=0$ is shown to exhibit a stable Fermi liquid phase for all $0<\xi<1$, although the quasiparticle weight becomes weaker as $\xi$ approaches unity. For $-1<\xi<0$, the value $\widetilde{Z}_k/Z_k^{\xi+1}$ is larger than one, implying a non-Fermi liquid like phase depending on the initial value of $Z_k$. The Fermi liquid ansatz for $\phi_0(\omega=0)=0$ is shown to break down for all $\xi>1$ and $\xi<-1$, as the effective quasiparticle weight becomes negative. Interestingly, if we allow the value of the $\xi=0$ retarded Green's function phase to shift from zero to some finite number, we may recover small pockets of Green's functions with powers $\xi>1$ where the effective quasiparticle weight remains non-zero and smaller than one. Physically, this means that Fermi liquid theory is possible for large non-trivial powers of the propagator, provided the imaginary part of the self-energy goes to zero at the same rate as the real part in the vicinity of the Fermi surface. From previous work by the author \cite{Heath1}, this implies that the imaginary part of the self energy itself goes as a power law $\Im \Sigma({\bf k},\,\omega)\sim \omega^\alpha$ with $\alpha<1$. Such self energy behavior has been noted in the pseudogap Bi$_2$Sr$_2$CaCu$_2$O$_{8+\delta}$, with NMR measurements in the pseudogap phase of the related cuprate Bi$_2$Sr$_{2-x}$La$_x$CuO$_{6+\delta}$ suggesting a Fermi liquid-like ground state \cite{Wen2,Zheng,Zheng2}. \added{Experimentally, a Landau-Fermi liquid like ground state has also been noted in the pseudogap phase of HgBa$_2$CuO$_{4+\delta}$ \cite{Mirzaei13,Barisic,Chan} and the high-field limit of underdoped YBa$_2$Cu$_3$O$_{6.5}$ \cite{Doiron} and YBa$_2$Cu$_4$O$_8$\cite{Proust16}. The calculation of this section therefore provides evidence that, for pseudogap phases with specific values of $\phi_0(\omega=0)$, a Landau-Fermi liquid ground state remains stable.}}

{The above discussion} leads us to conclude that {a stable Landau-Fermi liquid remains possible as long as $\xi>-1$; i.e., the branch-cut singularity resulting from the power-law nature of the propagator does not include the pole of the propagator itself. This statement shouldn't be that surprising; as stated before, branch-cut singularities have been shown to be present in stable Fermi liquids due to the particle-hole continuum \cite{Finkle}. A similar phenomenon is seen in lattices of Sachdev-Ye-Kitaev dots, where the propagator's branch cut singularity can be ignored in the low-temperature Fermi liquid phase \cite{Patel2}. Only when the branch cut engulfs the pole at $G({\bf k}_F,\,\omega=0)^{-1}=0$ is the Landau-Fermi liquid theory no longer possible, as then the quasiparticle picture breaks down entirely. What is surprising is that, by "tuning" the phase of the retarded Green's function at $\xi=0$, we have the possibility of a stable Fermi liquid for arbitrary values of $\xi$. As we will show in the next section, this results in non-trivial behavior of the fermionic degrees of freedom away from the Fermi surface at non-integer powers.}
%
%
%
\\\\
\noindent {\it III. The status of Luttinger's theorem in power-law liquids.--}
%
In its original form, Luttinger's theorem states that the volume of the Fermi surface remains {an invariant constant} in the presence of interactions \cite{Luttinger1,Luttinger2,Luttinger3}. {Originally developed perturbatively for a conventional Fermi liquid, Luttinger's theorem has since been shown to be a robust feature of gapless fermionic systems, haven been shown to be valid for a 1D Tomonaga-Luttinger liquid \cite{Krastan,Affleck}, the Kondo lattice \cite{Oshikawa}, and certain Mott insulators \cite{Heath1}. In the context of power-law liquids as considered in this work, the status of Luttinger's theorem has been a subject of much debate, with its fragility for powers $0<\xi<1$ already being noted \cite{Phillips4}.}

In a nutshell, Luttinger's theorem depends solely upon the two following sum rules \cite{Vignale,Bedell1,Phillips4,Heath1}:
\begin{subequations}
\begin{align}
&\frac{i}{2\pi}\int \frac{d^d {\bf k}}{(2\pi)^d}\oint_{\mathcal{C}}d\omega \frac{\partial}{\partial \omega}\log(G({\bf k},\,\omega))=\frac{N}{2V}\label{8a}
\end{align}

\begin{align}
&-i\int \frac{d^d {\bf k}}{(2\pi)^d} \oint_{\mathcal{C}}\frac{d\omega}{2\pi} \left\{G({\bf k},\,\omega)\frac{\partial}{\partial \omega}\Sigma({\bf k},\,\omega)
\right\}=0\label{8b}
\end{align}
\end{subequations}

\begin{figure*}[htp]
  \begin{center}
  \subfloat[]{\label{fig:2a}
      \hspace{0mm}\includegraphics[width=.65\columnwidth]{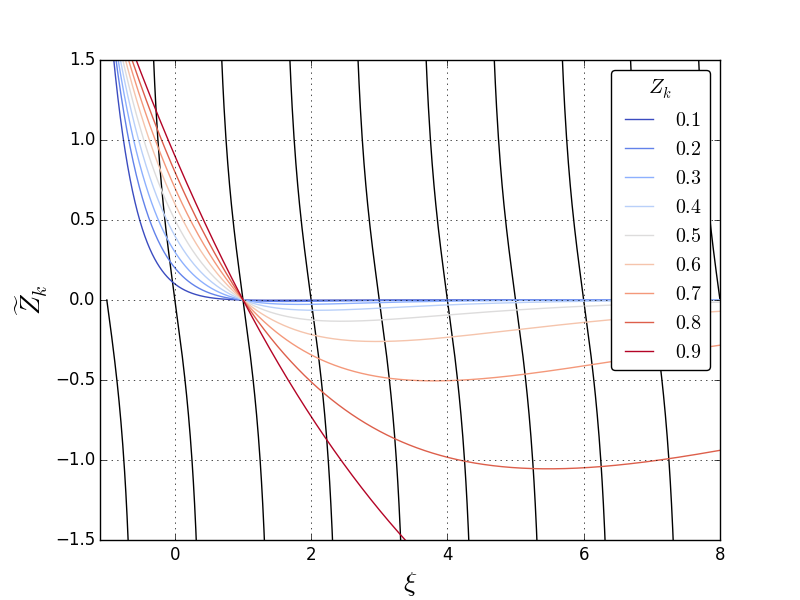}}
      \end{center}
\begin{center}
   \subfloat[]{\label{fig:2b}
        \hspace{0mm} \includegraphics[width=.65\columnwidth]{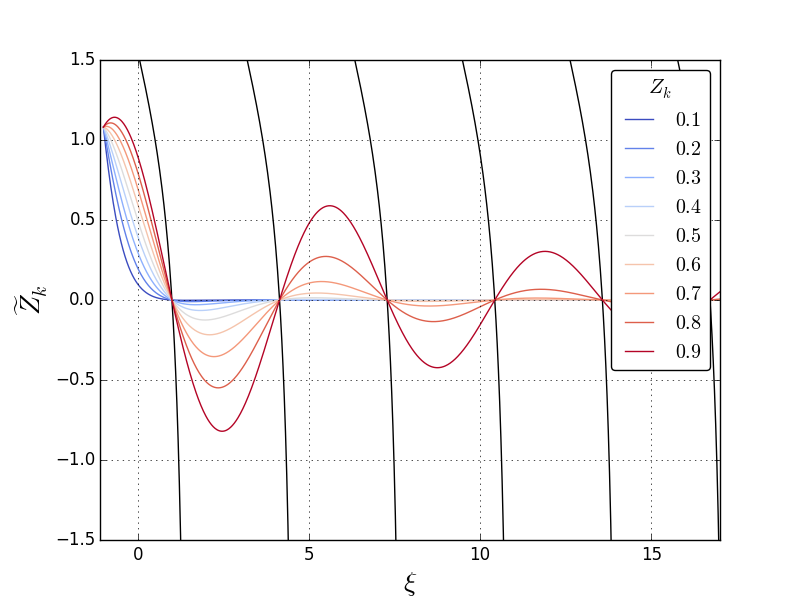}}
\end{center}
  \begin{center}
  \subfloat[]{\label{fig:2c}
      \hspace{0mm}\includegraphics[width=.65\columnwidth]{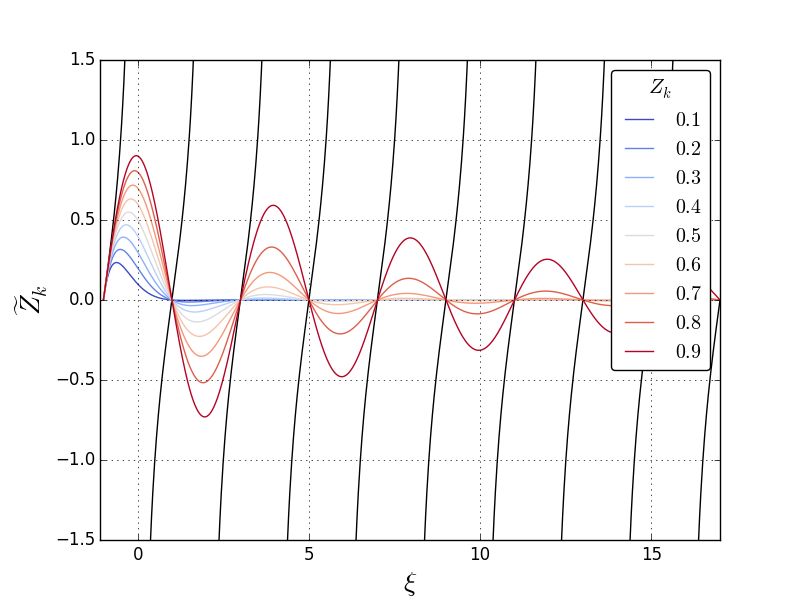}}
      \end{center}
   \caption{Values of the effective quasiparticle weight $\widetilde{Z}_k$ vs. $\xi$ (colour) plotted alongside the respective functions given in Eqn. \eqref{eq13} whose zeroes suggest the applicability of Luttinger's theorem (black). In Fig. \ref{fig:2a}, the value of the phase is taken as in a conventional Landau-Fermi liquid. For $\xi>1$ and $\xi\approx -1$, either $\widetilde{Z}_k>1$ or $\widetilde{Z}_k<0$, and Fermi liquid theory no longer applies. In Fig. \ref{fig:2b}, however, we take $\phi_0(0)$ to be some arbitrary value, leading to a weak (but finite) effective quasiparticle weight for larger values of $\xi$. Here I take the phase to be some arbitrary constant; to ensure agreement with the condition Eqn. \eqref{eq11}, only phases which yield zeros of $\tan[\xi\phi_0(\omega)]-\tan[\phi_0(\omega)]$ at integer values of $\xi$ lead to the validity of Eqn. \eqref{8a}. In Fig. \ref{fig:2c}, we see similar behavior as in Fig. \ref{fig:2a}, although this result is highly improbable as it violates causality. In all three cases, zeros of the functions defined in Eqn. \eqref{eq13} occur when $\widetilde{Z}_k=0$, leading us to conclude that Luttinger's theorem and Landau-Fermi liquid theory are always incompatible for power-law fluids.}\label{fig:2}
 
\end{figure*}

\noindent Eqn. \eqref{8b} requires the existence of a well-defined Luttinger-Ward functional, which is equivalent to $\Im \Sigma({\bf k},\,\omega)\sim \omega^\alpha$, with $\alpha>0$ \cite{Phillips6,Heath1}. In the case of strong self energy correlation, we can identify $\alpha\approx \xi+1$ \cite{Phillips4}, leading to $\xi=-1$ as the lower limit for a $\Phi$-derivable \cite{Luttinger2,Baym1,Cornwall} strongly correlated power-law liquid. Nevertheless, as long as we assume weak self energy correlation, this limit can be relaxed.

Eqn. \eqref{8a} can be recast in terms of the phase of the retarded Green's function, leading to the constraint that

\begin{align}
\frac{N}{2V}&=-\frac{1}{\pi}\int \frac{d^d {\bf k}}{(2\pi)^d } \left\{
\phi(0)-\phi(-\infty)
\right\}
\end{align}
Hence, for Luttinger's theorem to be obeyed in a generic fermionic system, we must enforce that $\tan[(\phi(0))]=\tan[(\phi(-\infty))]=0$.  

The phase $\phi_\xi(\omega)$ for non-zero $\xi$ can be {found by first writing down the imaginary and real parts of the power-law Green's function, in a similar manner as we did in Eqns. \eqref{A} and \eqref{ratio}:}

\begin{subequations}
\begin{equation}
{\Im G(k,\,\omega)\sim \Im \Sigma\frac{
\cos[\xi \phi_0(\omega)]-\cot[\phi_0(\omega)]\sin[\xi \phi_0(\omega)]
}{((x-\Re\Sigma({\bf k},\,\omega))^2+\Im\Sigma({\bf k},\,\omega)^2)^{\xi/2}
}}
\end{equation}

\begin{equation}
{\Re G_R \left({\bf k},\,\omega \right)\sim (x-\Re\Sigma ({\bf k},\,\omega)) \frac{
\cos[\xi \phi_0(\omega)]-\tan[\phi_0(\omega)]\sin[\xi \phi_0(\omega)]
}{((x-\Re\Sigma({\bf k},\,\omega))^2+\Im\Sigma({\bf k},\,\omega)^2)^{\xi/2}
}}
\end{equation}
\end{subequations}
{Combining these two results yields the equation connecting the phase of the power-law Green's function with that of the $\xi=0$ propagator:}

\begin{align}
\tan [\phi_\xi(\omega)]
&=\frac{
\tan[\phi_0(\omega)]\cos[\xi \phi_0(\omega)]-\sin[\xi \phi_0(\omega)]
}{
\cos[\xi\phi_0(\omega)]-\tan[\phi_0(\omega)]\sin[\xi\phi_0(\omega)]
}
\end{align}
The condition that $\phi_0(-\infty)=\pi$ is a direct consequence of causality, and hence a universal constraint on any fermionic system \cite{Tremblay,Heath1}. In the context of power-law Green's functions, we simply replace the condition of $-\infty$ with some UV cutoff $-\Delta$ which we assume to be reasonably large such that $\phi_0(-\Delta)=\pi$ to reasonable precision. Therefore, we can see that
\begin{align}
\tan[\phi_\xi(-\Delta)]=-\tan[\xi \pi]\label{eq11}
\end{align}
This is guaranteed to equal zero for integer values of the power $\xi$, which agrees with the result for $\xi=0$ when we take $\Delta\rightarrow \infty$. 

For the zero-frequency case, the validity of Luttinger's theorem depends upon the limiting value of $\tan[\phi_0(0)]$, as outlined below:

\begin{align}
&\tan [\phi_0(0)] =\begin{cases}
0,\,\, &-\tan [n\pi \xi]=0,\quad \xi\not=1\\
\textrm{const}
,\,\, &\tan \left[\xi \phi_0(0)\right]=\tan\left[\phi_0(0)\right],\quad \xi \not=1+\frac{n}{m+1/2}\\
\infty,\,\, &-\cot\left[\pi(m+1/2)\xi\right]=0,\quad \xi\not=1  
\end{cases}\label{eq13}
\end{align}

\noindent where $m,\,n\in\mathbb{Z}$. The first limit in the above is the common one encountered when $\xi=0$, and corresponds to an imaginary part of the self energy which goes to zero faster than the real part; i.e., $\Im\Sigma({\bf k},\,\omega)\sim \omega^\alpha$, $\alpha>1$ \cite{Heath1}. The second term corresponds to the real and imaginary portions of the self energy going to zero at the same rate, i.e. $\Im \Sigma({\bf k},\,\omega)\sim \omega^\alpha$, $0<\alpha<1$, while the final limit corresponds to the imaginary portion either going to a constant or going to zero slower than the real portion. As the final of these conditions violates causality, the first two limits are of greatest physical concern. The case when $\xi=1$ is special in that $\tan[\phi_0(0)]$ can be any arbitrary value for Luttinger's theorem to remain valid.

As the effective quasiparticle weight $\widetilde{Z}_k$ is also dependent on the phase $\phi_0(\omega=0)$, we can check the compatibility of the above relations with the existence of a well-defined Landau quasiparticle. In Fig. \ref{fig:2}, I plot  $\widetilde{Z}_k$ alongside the underlying conditions for Luttinger's Theorem given in Eqn. \eqref{eq13} vs. $\xi$. Note that, if we take values of $\phi_0(0)$ for the case of $|\tan[ \phi_0(0)]|\in \mathbb{R}_{>0}$ such that $\tan[\xi \phi_0(0)]-\tan[\phi_0(0)]=0$ only for $\xi\in \mathbb{Z}$, 
 then we can see that such a limit is compatible with the condition $\tan[\phi_\xi(-\Delta)]=0$; i.e., by Eqn. \eqref{eq11}, Luttinger's theorem is only valid at integer values of $\xi$. This agrees with the work of Limtragool et. al. \cite{Phillips4}, where Luttinger's theorem is strongly violated for $-1<\xi<1$, $\xi\not=0$. For the case $\xi=1$, Luttinger's theorem is always applicable, irrespective of the behavior of the self energy, although the effective quasiparticle weight will always be zero.

It is important to note that, for all cases examined in the above excluding $\tan[\phi_0(0)]=0$ with $\xi=0$, Luttinger's theorem is incompatible with a finite-valued effective quasiparticle weight. 
Although this result is surprising, it is not unheard of; the sharp peaks in underdoped samples of Bi$_2$Sr$_{2-x}$La$_x$CuO$_{6+\delta}$\cite{Meng1,Meng2} and the small Fermi surfaces in the underdoped phases of \added{HgBa$_2$CuO$_{4+\delta}$ \cite{Chan2}, YBa$_2$Cu$_4$O$_8$\cite{Proust16}, and} YBa$_2$Cu$_3$O$_{6.5}$\cite{Doiron} have led to the postulate that the non-superconducting pseudogap phase of the cuprate superconductors (in the absence of Fermi arcs) is an exotic variant of the traditional Fermi liquid where fermionic degrees of freedom deep in the Fermi surface are lost in the presence of strong interactions \cite{Wen2,Sachdev1,Punk,Sachdev2,Sachdev3}. {As stated previously, the Fermi liquid phase is stable for the pseudogap phase characterized by $\Im \Sigma({\bf k},\,\omega)\sim \omega^\alpha$ with $\alpha<1$, but only if the phase of the retarded Green's function at $\xi=0$ is non-trivial. Hence,} if such \added{an exotic} metallic state \added{in the pseudogap} is truly described by power-law Green's functions, the work presented above shows that \added{the incompatibility of Luttinger's theorem and Fermi liquid theory} is not a coincidence, but \added{is instead} a deep physical law that must be obeyed for any scale-invariant electronic propagator. 
\\\\
{\it IV. Discussion.--} 
{I have shown that Landau-Fermi liquid theory is possible in the presence of power-law scaling of the retarded Green's function as long as said scale invariance does not change the underlying topology of the Fermi surface. The power-law scaling can be recast in the form of a rescaled quasiparticle weight, which becomes zero for powers $\xi>1$ unless the phase of the retarded Green's function at unity power takes on non-trivial values. It is then shown that Luttinger's theorem breaks down for all cases of a stable Fermi liquid solution, excluding the conventional $\xi=0$ solution. This agrees with recent experiments on the pseudogap phase of cuprate superconductors, where a Fermi liquid-like phase is seen in the presence of possible power-law scaling.}


It is interesting to compare this result with the contradictory prediction that scale-invariance always leads to the breakdown of particle-like excitations in the IR \cite{Georgi,Georgi2}. Such an "unparticle" regime is usually characterized by the unparticle dimension $d_U$, which can be related to our coefficient $\xi$ via the relation
$
\xi=\frac{d-1}{2}-d_U
$,
where $d$ is the physical dimension. 
The conditions of unitarity constrain $d_U>\frac{d-1}{2}$ \cite{Minwalla,Phillips7}, leading to a power-law scaling of $-1<\xi<0$ being compatible with both the predictions of unparticle physics and a well-defined Landau quasiparticle in $d\ge 2$.
%
The existence of such a discrepancy can be justified in the context of quantum critical phenomena by recalling the effects of particle-hole asymmetry on the two-point time-dependent correlation function of the order parameter \cite{Ortiz}. 
{If quantum critical scaling induces a scale-invariant form of the Green's function as given in Eqn. \eqref{eq1}, then the system will experience severe particle-hole asymmetry either below or above the Fermi surface, and hence the emergent bosonic order parameter from the fermionic degrees of freedom (from Cooper pairing or otherwise) will be particle-hole asymmetric as well. }
Because the order parameter field for a system defined by Eqn. \eqref{eq1} is particle-hole asymmetric, 
the spatial correlation length will be finite, resulting in a finite scattering cross section and the possibility for dominant Fermi-liquid-like behavior.

Finally, the above might have important implications in the context of interacting, itinerant Majorana fermions, which have been shown to exhibit a finite discontinuity in the momentum distribution function while also displaying a small Fermi surface \cite{Baskaran,Heath2}. In the context of a Landau-Majorana liquid \cite{Heath4}, the presence of severe particle-hole asymmetry has been shown to lead to robust stability of the Landau quasiparticle state, which agrees with the work presented in this paper. It would be an interesting direction of future research to pursue the similarities between power-law liquids and Landau-Fermi-like Majorana liquids, and explore how such a theory can further elucidate the non-trivial physics of the cuprate pseudogap phase.
%
%
\\\\
\noindent {\it Acknowledgements.--}The author thanks Kevin Bedell, Krastan Blagoev, Matthew Gochan, and Kridsanaphong Limtragool for useful comments and suggestions. 

\bibliographystyle{iopart-num}

\bibliography{main}{}

\end{document}